\documentclass[twocolumn, noshowpacs, preprintnumbers, amsmath,amssymb,nopreprintnumbers]{revtex4}
\usepackage[colorlinks,urlcolor=blue]{hyperref}
\usepackage{graphicx}
\usepackage{lineno}
\usepackage{dcolumn}
\usepackage{bm}
\usepackage{appendix}
 \usepackage{verbatim}

\usepackage{breakurl}
\usepackage{amsmath}
\usepackage{enumerate}
\usepackage[final]{changes}     
\definechangesauthor[name={Author}, color=red]{Author}

\makeatletter
  \let\@font@info\@gobble
  \let\@font@warning\@gobble
\makeatother

\begin{document}

\title{Large deviations of the ballistic L{\'e}vy walk model}

\author{Wanli Wang}


\author{Marc H\"{o}ll}

\author{Eli Barkai}%
\affiliation{%
Department of Physics, Institute of Nanotechnology and Advanced Materials, Bar-Ilan University, Ramat-Gan
52900, Israel
\\
 }%


\date{\today}

\begin{abstract}
We study the ballistic L\'evy walk \added{stemming from an infinite mean traveling time between collision events.} Our study focuses on the density  of spreading particles all starting from a common origin,  which is limited by a `light' cone $-v_0 t<x<v_0 t$. In particular we study this
density close to its maximum in the vicinity of the `light' cone. The spreading density follows the Lamperti-arcsine law describing typical fluctuations far from the `light' cone. However this law blows up  in the vicinity of the `light' cone horizon which is nonphysical, in the sense that any finite time observation  will never diverge.  We claim that one can find two laws for the spatial density, the first one is the mentioned Lamperti-arcsine law describing the central part of the distribution and the second is an infinite density illustrating the dynamics for large $x$. 
We  identify the  relationship between a  large position and the longest  \added{traveling time} describing the single big jump principle.
From the renewal theory we find that the distribution of rare events of the position is related to the derivative of the average of the number of renewals at a short `time' using a rate formalism. 
%
%
%



\end{abstract}

\pacs{02. 50. -r,  05. 20. -y,  05. 40. -a }
%

%


\maketitle

\section{Introduction}\label{19lwsect1}

\begin{figure}[htb]
 \centering
 \includegraphics[width=8cm, height=6cm]{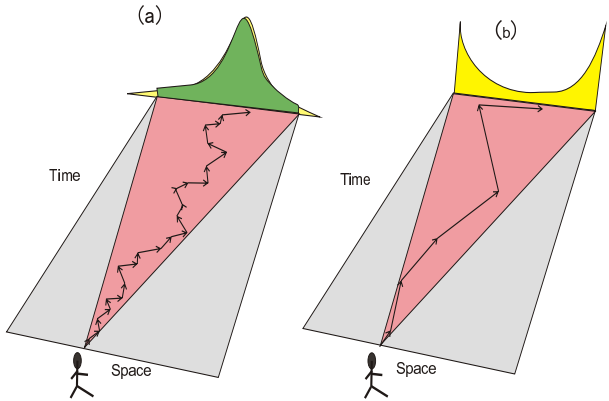}\\
 \caption{A test particle starting on the origin, has velocities $\pm v_0$. The traveling times are exponentially distributed (a) or 
power law distributed as for the ballistic L{\'e}vy walk (b). In these models we have a  ballistic cone marked by magenta and $-v_0 t<x(t)<v_0 t$.  The packet of spreading particles is described by the telegraph equation (a)
(solution in green) while the diffusion approximation (yellow in a) is invalid as it predicts particles exceeding the ballistic cone.  
For the ballistic  L{\'e}vy walk, panel  (b) the packet of particles is modeled by the Lamperti-arcsine  distribution, schematically presented in yellow. This solution diverges close to the ballistic cone, which as mentioned in the text is unphysical. In this work we explore the density close to the ballistic cone,
namely in the vicinity  of the maximum of the spreading packet, finding the deviations from the Lamperti-arcsine law. 
 }\label{myfig}
\end{figure}

\added{The diffusion equation describes the spreading of Brownian particles in a medium \cite{Einstein1905Uber,Smoluchowski1906Zur,Metzler2000random}. For a packet starting on the origin, the spreading  packet is Gaussian which is in agreement with the central limit theorem. Naively, this theory predicts a non-zero probability for a particle starting in say Jerusalem to be found in Tokyo in a split of a second. While this would be a rare event, it is in violation of causality. This problem was solved long time ago with the introduction of the telegraph equation \cite{Davydov1934Uravneniya,Goldstein1951diffusion,Joseph1989Heat,Masoliver1996Finite,Weiss2002Some,Bakunin2003Mysteries,Masoliver2017Continuous,Zaburdaev2015Levy}. Here, the particle in one dimension at position $x(t)$ with time $t$ is restricted to a `light'  or ballistic cone $-v_0t<x(t)<v_0t$ where $v_0>0$ is the typical speed (see Fig.~\ref{myfig}). The telegraph equation and its extensions to fractional telegraph equations are well studied \cite{Compte1997generalized,Eckstein1999mathematics,Cascaval2002Fractional,Qi2011Solutions,Fedotov2016Single,Awad2019On}. }

\added{Another well investigated problem is the L\'evy flight \cite{Zumofen1993Scale,Fogedby1994Levy,Shlesinger1995Levy}. In one dimension this random walk model deals with the sum of spatial displacements all drawn from a typically symmetric distribution which is fat tailed. The variance is infinite indicating that the L{\'e}vy type of central limit theorem is applicable. This process can be described by a fractional space diffusion equation $\partial P(x,t)/ \partial t = \nabla_x^\mu P(x,t)$ with $0<\mu< 2$ \cite{Metzler2000random,Kessler2012Theory}. Similar to the usual diffusion equation, the L\'evy flight and its corresponding fractional space diffusion equation suffer from the same deficiency. Here, small and large displacements are assumed all to take place in a unit of time. But this is nonphysical in most cases. In particular, the infinite mean squared displacement predicted by these theories  $\langle x^2 \rangle = \infty$ is non-realistic.}

\added{The solution to this problem is the introduction of the well-known and widely applicable L\'evy walk \cite{Shlesinger1982Random,Klafter1994Levy,Zaburdaev2011Perturbation,Zaburdaev2013Space,Ramos2004Levy,Zaburdaev2015Levy,Ariel2015Swarming,Magdziarz2016Explicit,Zaburdaev2016Superdiffusive,Fedotov2016Single,Marcin2017Aging,Fouxon2017Limit,Song2018Neuronal,Giona2019Age,Bologna2020Distribution,Extended2020Giona,Kanazawa2020Loopy}, where a finite velocity is introduced. Again, for particles starting on the origin we have a ballistic cone $-v_0t< x(t) < v_0t$. The mean squared displacement $\langle x^2(t) \rangle \le (v_0t)^2$ being  not faster than ballistic is in agreement with common sense physics.
Here, the main point of the L{\'e}vy walk model is the coupling between the walking time (or time of walk or traveling time) and the spatial displacement. Unlike  the L{\'e}vy flight, the stretch of the displacement is connected to its time cost, i.e., the walker has a finite velocity. }

\added{Here, we focus on the ballistic phase of  the  L{\'e}vy walk model' dynamics
and the density of spreading particles $P(x,t)$. In this case the Lamperti-arcsine \cite{Margolin2005Nonergodicity,Froemberg2015Asymptotic} distribution describes the shape of the distribution of propagating particles $P(x,t)$ in the long time limit (see below). In some sense, this distribution replaces the more familiar Gaussian and L\'evy distributions describing Brownian motion and L\'evy flight. The Lamperti-arcsine distribution has a U or W shape (see below). It means that $P(x,t)$ in this scaling limit diverges when $|x|\to v_0t$ (but always $|x| < v_0t$). The conclusion from this behavior is apparent. The divergence of the Lamperti-arcsine scaling solution is unphysical, in the sense that for any finite time we cannot obtain a blow up of the density, and this problem is cured here. 
In other words: the study of rare events is important. 
Hence, we set out to find the corrections and the accurate description of the L\'evy walk. Here the central part of $P(x,t)$ describes what we call typical or bulk  fluctuations while its behavior close to the ballistic cone is a rare event regime, to be defined more precisely later.  
To summarize, while the telegraph equation solves the nonphysical behaviour presented in the diffusion approximation and the L\'evy walk corrects the nonphysical nature of the L\'evy flight, we focus on the nonphysical blow up of Lamperti-arcsine
solutions of the ballistic L{\'e}vy walk. For schematics, see Fig. \ref{myfig}. }

\added{We emphasize that previous works in the field are technically  correct in the long-time limit $t\to\infty$. However, for large $t$ clear finite time effects are found here. And the rare event corrections are found when the particle distribution attains its maximum. In this sense, we are dealing with a vastly different case compared to the telegraph equation which provides $P(x,t)$ for large $x$ where the probability $P(x,t)$  is small at least for long times; see Fig.~\ref{myfig}. Our treatment of rare events of the ballistic L\'evy walk is based on the so called infinite covariant densities \cite{Aaronson1997introduction,Rebenshtok2014Infinite,Rebenshtok2014Non,Erez2017Large,Wang2018Renewal,Erez2019From,Wang2019Ergodic} and the big jump principle \cite{Cistjakov1964theorem,Alessandro2019Single,Wang2019Transport}. Both techniques are directly related (see below) and cure the nonphysical behaviour of Lamperti-arcsine solution in the vicinity of their maximum. This problem was already treated for the L\'evy walk with  finite mean traveling times using  several approaches  \cite{Fouxon2017Limit,Alessandro2019Single,Vezzani2020Rare}.}

The remainder of the manuscript is organized as follows. In Sec.~\ref{19LGS9}, we outline the L\'{e}vy walk model.   We study the difference between typical fluctuations and rare events, and compare them with simulations in Sec. \ref{19LGSlvyf104}.  In
Sec. \ref{18lwsec5}, we build the relationship between the position of the particle and the longest waiting time, exposing the big jump principle \cite{Alessandro2019Single} for the studied case.
The relation between rare events of the position and the averaged number, and the propagator are considered in  Secs. \ref{18lwsec5new} and \ref{18lwsec5finiewqv1}.
Finally, we conclude  with a discussion.

\section{Model}\label{19LGS9}
\subsection{Renewal process and L\'{e}vy walk model}\label{Moedel}
We first outline the main ingredients of the renewal process \cite{Feller1971introduction,Godreche2001Statistics, Godreche2015Statistics,Wang2018Renewal} and L\'{e}vy walk model.
The former  is defined as follows: Events happen at the random epochs of   time $t_1$, $t_2$, $\cdots$, $t_N$, $\cdots$, from some time origin $t=0$. Here we suppose time intervals $\tau_1=t_1$, $\tau_2=t_2-t_1$, $\cdots$, $\tau_N=t_N-t_{N-1}$, $\cdots$,  are  independent and identically distributed (IID) random variables with a common PDF $\phi(\tau)$. These $\tau_i$-s are called walking times (sometimes also flight times or traveling times).
Thus, the considered process is a renewal process. Given that the number of renewals during $(0,t)$ is $N$, i.e., $N=\max\{N,t_N\leq t\}$, the corresponding observation time $t$ is
\begin{equation}\label{2019lgsecA101}
  t=\sum_{j=1}^N\tau_j+B_t.
\end{equation}
Here $B_t$, defined by $t-t_N$, is the time interval between the time $t$ and the last event before $t$. When $t$ is fixed, our $N$ is a random variable.

We further consider the L\'{e}vy walk model in which  the directions of each step are introduced. The particles move continuously with a constant velocity $\pm v_0$ for a random time $\tau_1$ drawn from a PDF $\phi(\tau)$. Here the directions of particles, i.e., $+$ or $-$, are chosen randomly with equal probability. The corresponding displacement is $x_1=-v_0\tau_1$ ($x_1=v_0\tau_1$) on condition that the direction of the first step is negative (positive). We further generate another waiting time $\tau_2$ from $\phi(\tau)$ and the direction of the particle. Then the process is renewed. Here as mentioned $\tau_i$ are IID random variables with a common PDF $\phi(\tau)$. We are interested in the position of the particle at time $t$
\begin{equation}\label{2019lgsecA102}
  x(t)=\sum_{j=1}^N\chi_j+v_{j+1}B_t,
\end{equation}
where $\chi_j=\pm v_0\tau_j$~$(j=1,2,\cdots, N)$ are the displacement of $j$ step and $v_{j+1}B_t=\pm v_0B_t$ is the last displacement. \added{Notice that  $x$ is the position of the particle at time $t$ while $\chi$ is the displacement of the particle for a single step. Similar, $t$ describes the observation time of the process but $\tau$ is the time of walk or traveling time drawn from $\phi(\tau)$.}
Below we will show how to derive the distribution of $x(t)$. Clearly the particles starting on the origin are all within the `light' cone $-v_0 t \le x(t) \le v_0 t$.

\subsection{Propagator of L\'{e}vy walk}
Let us briefly recap the basic equations of the model considered in this paper.
For the velocity model under study, the particle moves continuously with a constant velocity and changes directions at random times \cite{Klafter1987Stochastic,Zaburdaev2015Levy}. Mathematically, the joint probability of the step's length $\chi$ and duration time $\tau$ is
\begin{equation}\label{19LGS9f101a1}
\phi(\chi,\tau)=\frac{1}{2}\phi(\tau)[\delta(\chi-v_0\tau)+\delta(\chi+v_0\tau)].
\end{equation}
The above equation describes the probability to move a distance $\chi$ in time $\tau$ with a single event and $\delta(|\chi|-v_0\tau)$  accounts for the space-time correlation.
The PDF of the particle's position at time $t$ is governed by \cite{Zumofen1993Scale}
\begin{equation}\label{19LGS9f103}
    Q(x,t)=\delta(t)\delta(x)+\int_0^t\int_{-\infty}^{\infty} Q(y,t^{'})\phi(x-y,t-t^{'})dydt^{'}
\end{equation}
and the PDF of the particle's position reads
\begin{equation}\label{19LGS9f104c}
    P(x,t)=\int_{-\infty}^{\infty}\int_{0}^{t}Q(y,t^{'})\Phi(x-y,t-t^{'}) dt^{'}dy,
\end{equation}
where $$\Phi(x, t)=\frac{1}{2}[\delta(x-v_0t)+\delta(x+v_0t)]\Psi(t)$$ is the probability of  moving a distance $x$ in time $t$ in a single motion during the last uncompleted step with $\Psi(t)=\int_t^{\infty}\phi(\tau)d\tau$, and $Q(x,t)$ is probability of
just arriving at $x$ at time $t$ after completing a step. In Eqs. (\ref{19LGS9f103}, \ref{19LGS9f104c}) we identify the convolution both in time and in space, hence the analysis proceeds with Laplace-Fourier transforms.
Combining Eqs. \eqref{19LGS9f103} and \eqref{19LGS9f104c} yields \cite{Klafter1987Stochastic,Zaburdaev2015Levy}
\begin{equation}\label{18lwsec3101}
\widetilde{\widehat{P}}(k,s)=\frac{\widehat{\Psi}(s+ikv_0)+\widehat{\Psi}(s-ikv_0)}{2-\left(\widehat{\phi}(s+ikv_0)+\widehat{\phi}(s-ikv_0)\right)},
\end{equation}
where $\widetilde{\widehat{P}}(k,s)$ is the Fourier $x\to k$ and Laplace $t\to s$ transforms of $P(x,t)$. Such equations  are known as  Montroll-Weiss equations, they are not generally easy to invert, and hence later we turn to the asymptotic analysis.

\subsection{Three types of distributions of  waiting times}
In our analysis, we use the power law distribution of the times \added{of walk} capturing a  heavy tail \cite{Peter1990Reaction,Bouchaud1990Anomalous,Metzler2000random,Fernando2009Beyond,Tameem2017Temperature}
\begin{equation}\label{19lwsect1sde}
\phi(\tau)\sim \tau^{-\alpha-1}
\end{equation}
with $0<\alpha<1$ for large $\tau$. In Laplace space, from  the Tauberian theorem \cite{Feller1971introduction} and Eq.~\eqref{19lwsect1sde} we have
\begin{equation}\label{2019lgsec102}
  \widehat{\phi}(s)\sim 1-b_\alpha s^\alpha, s\to 0
\end{equation}
with $0<\alpha<1$.
Here $b_\alpha$ is a constant determined by the details of $\phi(\tau)$. In this paper we denote $\widehat{\phi}(s)$  as the Laplace transform of $\phi(\tau)$ and $s$ is conjugate to $\tau$. We  have $\widehat{\phi}(0)=1$, since $\phi(\tau)$ is a normalized density.  Below we consider the exact forms of three types of waiting time PDFs with the same heavy-tails, which will be used to show the features of typical fluctuations and rare fluctuations.
\subsubsection{Pareto distribution}
Our first example is called the Pareto distribution \cite{Klafter2011First}. It is defined as follows
\begin{equation}\label{2019lgsec101}
\phi(\tau)=\left\{
          \begin{split}
            &0, & \hbox{$\tau\leq\tau_0$;} \\
            &\alpha\frac{\tau_0^\alpha}{\tau^{1+\alpha}}, & \hbox{$\tau>\tau_0$.}
          \end{split}
        \right.
\end{equation}
When $0<\alpha<1$, the first moment of $\tau$ is divergent. Note that 
for Eq.~\eqref{2019lgsec101}, we have $b_\alpha=\tau_0^\alpha|\Gamma(1-\alpha)|$ according to Tauberian theorem.

\subsubsection{One-sided L{\'e}vy distribution}
We further introduce the one-sided L{\'e}vy PDF $\ell_\alpha(\tau)$. In Laplace space, $\ell_\alpha(\tau)$ has a simple form
\begin{equation}\label{ldeq32011}
 \int_0^\infty \exp(-s\tau)\ell_\alpha(\tau)d\tau=\exp(-s^\alpha)
\end{equation}
and the small $s$ expansion is  $\widehat{\phi}(s)\sim 1-s^\alpha$  with $0<\alpha<1$. Let us first consider the special case of $\alpha=1/2$, i.e.,
\begin{equation}\label{ldeq32011levy}
  \ell_{1/2}(\tau)=\frac{1}{2\sqrt{\pi}}\tau^{-\frac{3}{2}}\exp\left(-\frac{1}{4\tau}\right), {\rm for}~~\tau>0.
\end{equation}
We see from Eq.~\eqref{ldeq32011levy}  that $\ell_{1/2}(\tau)\to 0$ for $\tau\to 0$.
However, for any small and finite $\tau$, $\ell_{\alpha}(\tau)\neq 0$, which is obviously different from Eq.~\eqref{2019lgsec101}.
%

\subsubsection{Mittag-Leffler distribution}
Another density of the walking time is the Mittag-Leffler PDF \cite{Podlubny1999Fractional,Kozubowski2001Fractional,Vainstein2006Non,Oliveira2019Anomalous}, i.e.,
\begin{equation}\label{ldbeq102h}
\phi(\tau)=\tau^{\alpha-1}E_{\alpha,\alpha}(-\tau^\alpha),0<\alpha<1
\end{equation}
with $E_{\alpha,\alpha}(\cdot)$ being the Mittag-Leffler function  defined by
\begin{equation}\label{ldbeq102u}
E_{\alpha,\beta}(z)=\sum_{n=0}^{\infty}\frac{z^n}{\Gamma(\alpha n+\beta)}.
\end{equation}
In Laplace space, $\widehat{\phi}(s)$ has the specific form
 \begin{equation}\label{ldbeq102hi}
\widehat{\phi}(s)=\frac{1}{1+s^\alpha}.
 \end{equation}
The Mittag-Leffler distribution is a geometric stable distribution \cite{Kozubowski2001Fractional}. When $\tau\to 0$, we have $\phi(\tau)\propto \tau^{\alpha-1}\to \infty$.

\subsection{Rare events versus typical fluctuations}



\begin{figure}[htb]
 \centering
 \includegraphics[width=8cm, height=6cm]{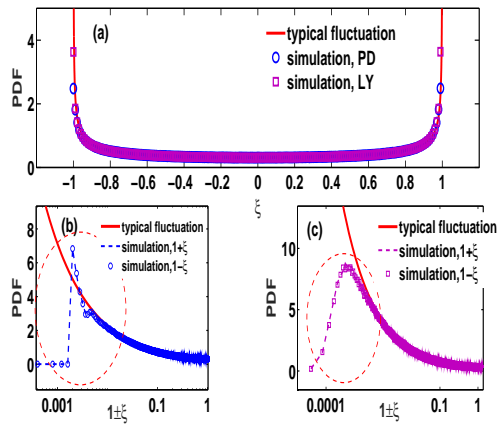}\\
 \caption{ The PDF of the position for L\'{e}vy walk model with $\alpha=1/2$ and $\xi=x/(v_0t)$. As expected the distribution of the position of the velocity model is symmetrical with respect to $\xi=0$. Note that in Figs. (b) and (c) we plot the PDFs versus  $1\pm \xi$ using the Pareto distribution Eq.~\eqref{2019lgsec101}  and the one-sided L\'{e}vy distribution Eq.~\eqref{ldeq32011levy} for the traveling time PDF. Clearly, when $1-|\xi|\to 0$, both sides  of the distribution of the position have deviation from the red solid line showing the typical fluctuations. The later are described by the arcsine law Eq. \eqref{18lwsec3101a4ooo}. Here `LY' and `PD' denote the one-sided L{\'e}vy distribution Eq.~\eqref{ldeq32011} and the Pareto distribution Eq.~\eqref{2019lgsec101}, respectively. In our simulations, we use $t=1000$, $\tau_0=1$, and $10^7$ realizations. In this manuscript we analyse the rare fluctuations circled in red. }\label{PreImage}
\end{figure}

For $0<\alpha<1$, the average of the waiting time diverges, which leads to  ballistic-diffusion \cite{Klafter1994Levy}, namely $\langle x^2(t)\rangle \sim (v_0)^2(1-\alpha) t^2$ with $t\to\infty$. \added{The infinite mean traveling time is responsible for the ballistic L{\'e}vy walk characterized by 
the ballistic front.
Note that also Generalised Langevin equation, under certain conditions, yields
ballistic diffusion, however since the noise in this equation is assumed Gaussian both the typical and rare events  are normal \cite{Vainstein2006Non,Oliveira2019Anomalous}.} 
In \cite{Froemberg2015Asymptotic}, the typical fluctuations were discussed in detail, i.e., the position $x$ is of the order of $t$.
When $\alpha=1/2$, the distribution of the position  follows the arcsine law \cite{Lamperti1958occupation,Froemberg2015Asymptotic}
\begin{equation}\label{18lwsec3101a4ooo}
 P_{\xi}(\xi)\sim \frac{1}{\pi\sqrt{1-\xi^2}}
\end{equation}
with \added{$-1<\xi=x/v_0t<1$}.  Clearly, the arcsine law works very well for the central part of the distribution of the position \added{both for one-sided L{\'e}vy distribution and Pareto distribution;} see the red solid line in Fig.~\ref{PreImage} (a). Due to the variable transformation, $P_\xi(\xi)$ and $P(x,t)$ are related by $P_\xi(\xi)=v_0tP(x=v_0t \xi,t)$.
While, when $|\xi|\to 1$ or $|x|\to \pm v_0t$, the typical fluctuations Eq.~\eqref{18lwsec3101a4ooo} blow up which is nonphysical at least for a finite time $t$.  This drawback of the arcsine law, i.e., the nonphysical  divergence at $x$ in the vicinity of the ballistic cone, is circumvented  in this paper when   a second type of scaling of the density is considered. See the data circled in red on the bottom panel of Fig.~\ref{PreImage}. It implies that  under certain conditions the density of the position is characterized by two scaling laws. The first one is the mentioned normalized arcsine law Eq.~\eqref{18lwsec3101a4ooo} \added{ describing the scaling when $|x|\propto v_0t$ but $|x|\not\simeq v_0t$. Note that the central limit theorem (Lamperti-arcsine form)
describes the central part of the distribution, but
for finite though large $t$ it does not describe the rare fluctuations,
i.e., large $x$ behavior.} \added{An important feature of the typical fluctuations is that its behavior is universal which is only determined by the far tail of the waiting time PDF, namely, Eqs.~(\ref{2019lgsec101}, \ref{ldeq32011})  and \eqref{ldbeq102h}, are not
vitally important for the typical fluctuations on condition that they have the same heavy-tails governed by the index $\alpha$.}

The second scaling corresponds to the non-normalized state \added{showing the behavior of the density for  $x\simeq v_0t$}, which is  described by infinite densities \cite{Aaronson1997introduction,Rebenshtok2014Infinite,Rebenshtok2014Non,Erez2017Large,Wang2018Renewal,Erez2019From,Wang2019Ergodic}. Here our aim is to find the statistics of rare fluctuations $|x|\approx v_0t$, where the detailed structure of the waiting time PDF is of importance.

\section{ Results for L\'{e}vy walk}\label{19LGSlvyf104}

\subsection{Bulk fluctuations}
First, we focus on the typical fluctuations, namely the case $|x|\propto v_0t$ and both are large,  implying that $s$ and $k$ are small and comparable. \added{Inserting Eq.~\eqref{2019lgsec102} into Eq.~\eqref{18lwsec3101}, we get}
\begin{equation*}
\widetilde{\widehat{P}}(k,s)\sim \frac{1}{s}\mathcal{G}\left(\frac{k}{s}\right)
\end{equation*}
with
$$\mathcal{G}(y)=\frac{\left(1+iv_0y\right)^{\alpha-1}+\left(1-iv_0y\right)^{\alpha-1}}{\left(1+iv_0y\right)^{\alpha}+\left(1-iv_0y\right)^{\alpha}}.$$
\added{It can be seen that the above equation is normalized since $\widetilde{\widehat{P}}(k\to 0,s)\sim 1/s$.
The inverse Fourier-Laplace transform for the expression $\mathcal{G}(y=k/s)/s$ can be performed exactly following the methods given in Refs.~\cite{Godreche2001Statistics,Froemberg2015Asymptotic}, which yields} the description of what we call bulk or typical fluctuations \cite{Margolin2005Nonergodicity,Froemberg2015Asymptotic,Magdziarz2016Explicit}
\begin{equation}\label{18lwsec3101a4}
\begin{split}
 P_{\xi}(\xi) & \sim \frac{\sin(\pi\alpha)}{\pi}\times \\
    & \frac{|1-\xi|^{\alpha}|1+\xi|^{\alpha-1}+|1+\xi|^{\alpha}|1-\xi|^{\alpha-1}}{|1-\xi|^{2\alpha}+|1+\xi|^{2\alpha}+2\cos(\pi\alpha)|1-\xi|^\alpha|1+\xi|^{\alpha}}
\end{split}
\end{equation}
with the scaling form $\xi=x/(v_0t)$. Here as usual, the subscript $\xi$ means that $P(\cdot)$ is the corresponding PDF of $\xi$.
The propagator Eq.~\eqref{18lwsec3101a4} is called  the Lamperti distribution \cite{Lamperti1958occupation}. Here $-1\leq\xi\leq 1$ since $-v_0t \leq x\leq v_0t$, namely there exists a finite `light' cone in which we may find the particle. The second moment of the position is $\langle x^2(t)\rangle\sim (1-\alpha) (v_0t)^2$, which corresponds to a ballistic behavior \cite{Klafter1994Levy,Froemberg2015Asymptotic}.
When $\alpha=1/2$, Eq.~\eqref{18lwsec3101a4} reduces to Eq.~\eqref{18lwsec3101a4ooo} which is plotted in Fig. \ref{PreImage}, and as expected describes well the central part of the packet of spreading particles.  A related expression is
\begin{equation}\label{18lwsec3101a5}
 P_{\epsilon}(\epsilon,t)\sim \frac{1}{\pi\sqrt{\epsilon(2v_0t-\epsilon)}}
\end{equation}
with \added{$0<\epsilon=v_0t-x<2v_0t$}
which is plotted by the dashed line in Fig.~\ref{LevywalkVeolicity05Large}. There when $\epsilon$ is small, we identify the deviations from the arcsine law.


\subsection{Rare fluctuations}
\begin{figure}[htb]
 \centering
 \includegraphics[width=8cm, height=6cm]{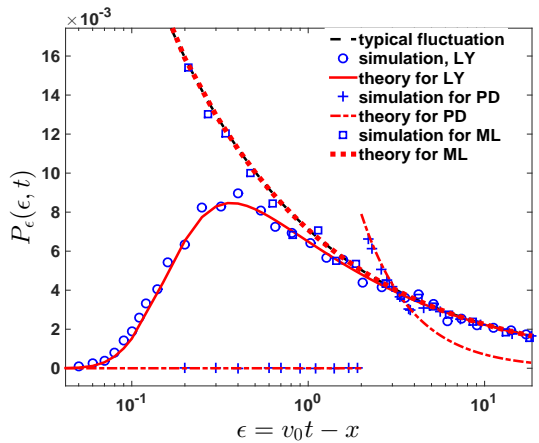}\\
 \caption{The behavior of $P_{\epsilon}(\epsilon,t)$ for small $\epsilon$ with $\epsilon=v_0t-x$. The full line [Eq.~\eqref{18lwsec3105}], the dash-dotted one [Eq.~\eqref{18lwsec310600}], and the dotted one [Eq.~\eqref{18lwsec3106f45}] describing the rare events are the theoretical predictions  with different waiting time distributions showing  different behaviors of rare fluctuations. The dashed  line Eq.~\eqref{18lwsec3101a4} is the  Lamperti distribution, which illustrates the PDF when both $x$ and $t$ are of the same order and comparable. The symbols are the simulation results  obtained by averaging $10^7$ trajectories of the particles with $\alpha=1/2$.
 Here `ML'  denotes the Mittag-Leffler distribution Eq.~\eqref{ldbeq102h}.
}\label{LevywalkVeolicity05Large}
\end{figure}

We consider the case of $x\to v_0t$  using the random variable $\epsilon=v_0t-x$ where $\epsilon$ is small. In Fourier-Laplace spaces, the  density of $\epsilon$ becomes
\begin{equation}\label{18lwsec3102}
\begin{split}
\widetilde{\widehat{P}}{\epsilon}(k_\epsilon,s)=\widetilde{\widehat{P}}(-k_\epsilon,s-ik_\epsilon v_0).
\end{split}
\end{equation}
Here $k_\epsilon$ is the Fourier pair of the shifted position $\epsilon$.
Utilizing  Eqs.~\eqref{18lwsec3101} and \eqref{18lwsec3102}, we get
\begin{equation}\label{18lwsec3103}
 \widetilde{\widehat{P}}_{\epsilon}(k_\epsilon,s)=\frac{\widehat{\Psi}(s-2ik_\epsilon v_0)+\widehat{\Psi}(s)}{2-[\widehat{\phi}(s-2ik_\epsilon v_0)+\widehat{\phi}(s)]}.
\end{equation}
We are interested in  analyzing the behavior of the position in the long time regime ($s\to 0$), where $\epsilon$ and $t$ are sufficient small and large, respectively. Using Eq.~\eqref{2019lgsec102}, Eq.~\eqref{18lwsec3103} reduces to a  simple expression
\begin{equation}\label{18lwsec3104}
\widetilde{\widehat{P}}_{\epsilon}(k_\epsilon,s)\sim \frac{\widehat{\Psi}(-2ik_\epsilon v_0)+b_\alpha s^{\alpha-1}}{1-\widehat{\phi}(-2ik_\epsilon v_0)}.
\end{equation}
Note that the inverse Laplace transform of $\widehat{\Psi}(-2ik_\epsilon v_0)/(1-\widehat{\phi}(-2ik_\epsilon v_0))$ gives a delta function $\delta(t)$ which is ignored and not related to our long time behavior. \added{Here, we stress that Eq.~\eqref{18lwsec3104} is valid in the limit of $s\to 0$ and $k_\epsilon\to \infty$. In other words, Eq.~\eqref{18lwsec3104} can not yield an effective prediction for $k_\epsilon\to 0$. See further discussion below.}
Taking  the inverse Laplace-Fourier transform of Eq.~\eqref{18lwsec3104} gives the main result of this section
\begin{equation}\label{18lwsec3105}
  \frac{t^{\alpha}\Gamma(1-\alpha) }{  b_\alpha } P_{\epsilon}(\epsilon,t)\sim \mathcal{I}(\epsilon)
\end{equation}
with
\begin{equation}\label{18lwsec3105a1}
\mathcal{I}(\epsilon)=\mathcal{F}^{-1}_{\epsilon}\left[\frac{1}{1-\widehat{\phi}(-2ik_\epsilon v_0)}\right].
\end{equation}
Here we used the fact that $s^{\alpha-1}$ and $t^{-\alpha}/\Gamma(1-\alpha)$ are Laplace pairs. \added{As mentioned before the time-dependent Eq.~\eqref{18lwsec3105} is valid in the limit of $t\to \infty$.}
\added{ The theoretical prediction} Eq.~\eqref{18lwsec3105} with $\phi(\tau)$ being the  one-sided L\'{e}vy distribution is plotted in Fig.~\ref{LevywalkVeolicity05Large} by using the numerical inverse Fourier transform. The comparison to numerical simulation is excellent, while the arcsine law completely fails to describe the observed behavior.
It can be seen that  $\mathcal{I}(\epsilon)$ given in Eq. (\ref{18lwsec3105a1}) is
an infinite density since $\widetilde{\widehat{P}}_{\epsilon}(k_\epsilon=0,s)\neq 1/s$, namely the  $\mathcal{I}(\epsilon)$ is not normalised,
which is hardly surprising since it is obtained from a normalised density multiplied by $t^\alpha$ hence the area under the left hand side of Eq.~\eqref{18lwsec3105} is obviously diverging.

Let us consider three examples:

i) For the  Pareto distribution, we can not invert Eq. \eqref{18lwsec3105} exactly, however we may invert it numerically. While, there is a simply way by  considering the limit $\epsilon\to 0$. Using $1/(1-\widehat{\phi}(-2ik_\epsilon v_0))\sim 1+\widehat{\phi}(-2ik_\epsilon v_0)+\widehat{\phi}(-2ik_\epsilon v_0)^2$ and taking the inverse Fourier transform gives
\begin{equation}\label{18lwsec3105a1d}    
\begin{split}
\mathcal{I}(\epsilon)\sim  & \delta(\epsilon)+\frac{1}{2v_0}\phi\left(\frac{\epsilon}{2v_0}\right)+\int_0^{\frac{\epsilon}{2v_0}}\frac{\phi(y)}{2v_0}\phi\left( \frac{\epsilon}{2v_0}-y\right)dy.
\end{split}
\end{equation}
If we are only interested in the behavior of $x\to v_0t$, the second term works perfectly and Eq.~\eqref{18lwsec3105a1d} reduces to
\begin{equation}\label{18lwsec310600}
 \mathcal{I}(\epsilon)\sim 
\frac{1}{ 2v_0}\phi\left(\frac{\epsilon}{2v_0}\right)
\end{equation}
with $\epsilon \neq 0$.
Note that Eq.~\eqref{18lwsec310600} in the limit $\epsilon\to 0$  is valid for a large range of PDFs, for example the mentioned Mittag-Leffler and the one-sided L\'{e}vy distributions.

ii) For the one-sided L\'{e}vy distribution, we use the geometric series $1/(1-\widehat{\phi}(-2ik_\epsilon v_0))=\sum_{n=0}^\infty \widehat{\phi}^n(-2ik_\epsilon v_0)$ and get by inversion $k_\epsilon\to\epsilon$
\begin{equation}\label{18lwsec3106}
\mathcal{I}(\epsilon)=\delta(\epsilon) + \sum_{n=0}^\infty\frac{1}{(2nv_0)^{1/\alpha}}L_{\alpha}\left(\frac{\epsilon}{(2nv_0)^{1/\alpha}}\right),
\end{equation}
where $L_{\alpha}(x)$ is the L\'{e}vy  PDF \cite{Metzler2000random}, defined by
\begin{equation}\label{deflevystable}
L_{\alpha}(x)=\frac{1}{2\pi}\int_{-\infty}^{\infty}\exp(-ikx)\exp[(ik)^\alpha]{\rm d}k.
\end{equation}
When $\epsilon=0$ or $x=v_0t$, only the function $\delta(\epsilon)$ is of importance. As expected, Eq.~\eqref{18lwsec3105} reduces to the survival probability, describing the probability of moving in the same direction for the whole observation time $t$. On the contrary, the function $\delta(\epsilon)$ loses its role for $\epsilon\neq 0$ or $x\neq v_0t$.

iii) For the Mittag-Leffler distribution, using Eq.~\eqref{ldbeq102hi}, it is easy to show
\begin{equation}\label{18lwsec3106f45}
\mathcal{I}(\epsilon)=\delta(\epsilon)+\frac{\epsilon^{\alpha-1}}{(2v_0)^{\alpha}\Gamma(\alpha)}.
\end{equation}
Utilizing Eqs.~\eqref{18lwsec3106f45} and \eqref{18lwsec3105}, as $t\to \infty$, we have
\begin{equation}\label{18lwsecverl3108}
P_{\xi}(\xi,t)\sim \frac{b_\alpha}{ t^{\alpha}\Gamma(1-\alpha)}\delta(1-\xi)+\frac{\sin(\pi\alpha)}{2^\alpha \pi}(1-\xi)^{\alpha-1}
\end{equation}
according to the relationship $t^{\alpha-1}E_{\alpha,\alpha}(-t^\alpha)\sim -1/(\Gamma(-\alpha)t^{\alpha+1})$. It indicates that  Eq.~\eqref{18lwsecverl3108}  agrees with the far tail of the Lamperti distribution Eq.~\eqref{18lwsec3101a4} in the limit of $\xi \to 1$. Namely, Eq.~\eqref{18lwsecverl3108} exhibits a unique behavior that the rare events are described by the same theory as the typical fluctuations; see Fig.~\ref{LevywalkVeolicity05Large}.
Besides, an interesting feature of $P_{\epsilon}(\epsilon,t)$  is exclusively exhibited by the rare events analysis, i.e., a discrete probability describing the survival probability of the particles $\Psi(t)\sim t^{-\alpha}$ is found. \added{ For the Mittag-Leffler distribution, it is easy to check that Eq.~\eqref{18lwsec3105} is not normalised. To be more exactly, the divergence happens at $\epsilon\to \infty$.}

\begin{figure}[htb]
 \centering
 \includegraphics[width=8cm, height=6cm]{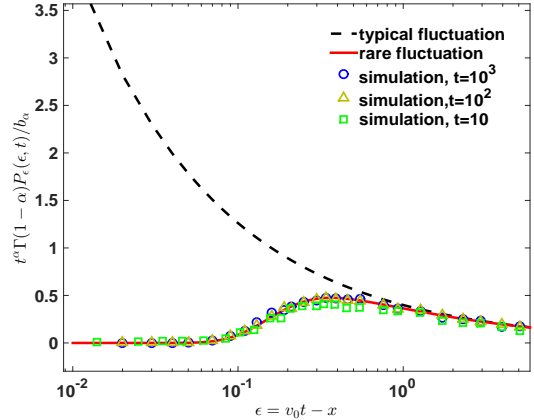}\\
 \caption{ The PDF of $\epsilon=v_0t-x$ multiplied by $t^{\alpha}\Gamma(1-\alpha)$ versus $\epsilon$ for a model where the travel time PDF $\phi(\tau)$ is the one-sided L{\'e}vy distribution. The solid line is the analytical solutions  $\mathcal{I}(\epsilon)$ [Eq.~\eqref{18lwsec3105a1}] obtained by the numerical inverse Fourier transform and the symbols are simulations with different $t$, namely, $t=10^3$, $t=10^2$, and $t=10$. Other parameters are the same as in Fig.~\ref{LevywalkVeolicity05Large}.
}\label{infinitedensityVeloLarge05}
\end{figure}

In Fig.~\ref{LevywalkVeolicity05Large},  we show the propagator corresponding to Eq. \eqref{18lwsec3105} in the scaling form. For $\alpha=1/2$, the calculated $P_{\xi}(\xi)$, Eq.~\eqref{18lwsec3101a4}, follows reasonably the arcsine law and this is only valid for the central part of the distribution of the position, namely $x\propto v_0t$ but $x \not\approx v_0t$. As the figure shows, it is difficult to find the difference between the typical fluctuations and the theoretical result with the Mittag-Leffler waiting time statistics, while, if $x\to v_0t$ and the waiting time follows the Pareto or the one-sided L{\'e}vy distributions, deviations from Eq.~\eqref{18lwsec3101a5} are clearly presented.
When $x\to -v_0t$, the rare events of $x$ can be obtained by using the symmetry property of the density and here we did not discuss this in detail. In Fig.~\ref{infinitedensityVeloLarge05}, the scaling form $\mathcal{I}(\epsilon)$ is exhibited with different observation time $t$ to show the properties of the infinite density. Clearly, for small $\epsilon$, $\mathcal{I}(\epsilon)$ is independent of the observation time $t$  and its shape does not change.

%
%
%

\section{Relation between the position and the longest waiting time}\label{18lwsec5}
\added{ An interesting problem is  the relation between
large $x$ behavior of the distribution  $P(x,t)$ and the distribution of the longest traveling time in the renewal process.
This problem is related to extreme value statistics \cite{Gumbel2004Statistics,Albeverio2006Extreme,Majumdar2009Large,Lefevere2011Large,Tsirelson2013uniform,Krapivsky2014Large,Chaitra2014Universal,Mauro2016Large,Hartich2019Extreme,Marc2019}. Extreme events are natural phenomena and  play an important role in our life. Thus, it is important to study  how these rare events are related to  other observables. Here, we wish to establish a connection between the longest  time of walk and the position of the particle. Such relations
are based on the well-known big jump principle  \cite{Cistjakov1964theorem,Alessandro2019Single}. Mathematically, when $\tau_1,\tau_2, \cdots,\tau_N$ are IID random variables with a sub-exponential tail, the single big jump principle is}
\begin{equation}\label{18lwsec5eq03ddf1}
\begin{split}
{\rm Prob}(\tau_1+\tau_2&+\cdots+\tau_N\geq z)\\
 &={\rm Prob}(\max\{\tau_1,\tau_2,\dots,\tau_N\}\geq z)
\end{split}
\end{equation}
when $z$ is large.  \added{Notice that this relationship Eq.~\eqref{18lwsec5eq03ddf1} is valid for any integer $N\ge 1$.} See related works where this principle was applied in physical systems
\cite{Thomas2013Precise,Buraczewski2013Large,Wang2019Transport}. However here we have a constraint on the time of walk, namely they all sum up to the fixed measurement time $t$. In other words, the traveling times are non IID and Eq.~\eqref{18lwsec5eq03ddf1} is not directly relevant to our study. 
\added{We will  see that the big jump principle for the ballistic motion is different in  comparison to the previously studied case when $\alpha>1$} \cite{Alessandro2019Single}.

\begin{figure}[htb]
 \centering
 \includegraphics[width=8cm, height=5.5cm]{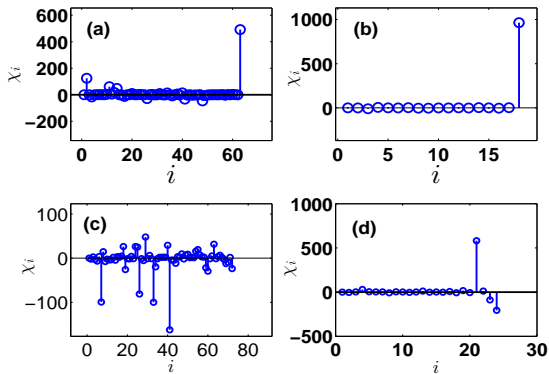}\\
 \caption{Step size $\chi_i$ for the velocity model when the PDF of travel times is  the one-sided L{\'e}vy distribution and $\alpha=0.5$. The observation time $t$ is $1000$ and $i=1$, $2$, $\cdots$, correspond to the first, second, $\cdots$, waiting time of the L{\'e}vy walk model, respectively.  The step length of each step denoted by $\chi_i$ is $\chi_i=\pm v_0\tau_i$ with \added{$\tau_i=t_i-t_{i-1}$} being  the $i$-th time of walk. Note that the directions of the particles are either $+$ or $-$ chosen randomly with equal probability. We see that one displacement is dominating the land-scope, this is the biggest jump. In this section we mainly consider the relation between the largest position and the longest waiting time.
}\label{tauMAXx}
\end{figure}

For the well-known IID case, $N$ in  Eq.~\eqref{18lwsec5eq03ddf1} is  fixed.
In our model, $N$ is a random variable since $\tau_1+\tau_2,\cdots,+\tau_N+B_t=t$; see Fig.~\ref{tauMAXx}.
This
constraint also implies that we have correlations in the process, though they stem from a renewal  process, and hence still can be analysed.  Here we must distinguish between two types of rare events. We will focus on $x$ being large but strictly $x<v_0 t$. Then in Sec.\ref{18lwsec5finiewqv1} we will treat $x= v_0 t$. The latter gives a delta function contribution to $P(x,t)$.

\begin{figure}[htb]
 \centering
 \includegraphics[width=8cm, height=6cm]{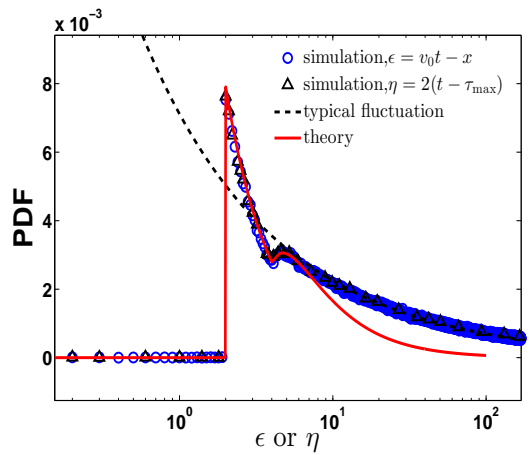}\\
 \caption{Densities of $\epsilon=v_0t-x$  (L\'{e}vy walk model) and $\eta=2v_0(t-\tau_{\max})$ (renewal process) with the Pareto distribution Eq.~\eqref{2019lgsec101}. The  red solid line Eq.~\eqref{18lwsec3105a1d} illustrating the rare events is the theoretical prediction and the symbols are simulations. Here we use  the same parameters as in a previous Fig. \ref{LevywalkVeolicity05Large}.
}\label{PositionVsTauMaxSimulation}
\end{figure}

\begin{figure}[htb]
 \centering
 \includegraphics[width=8cm, height=5.5cm]{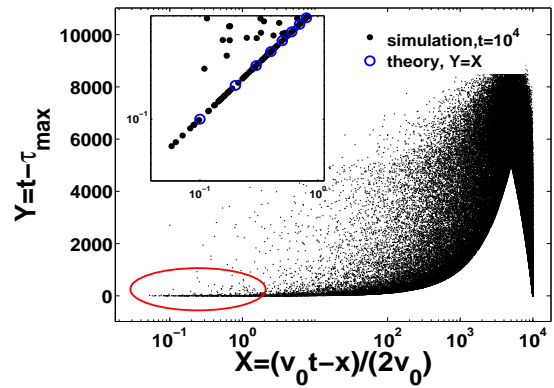}\\
 \caption{A correlation plot between $X=(v_0t-x)/(2v_0)$ and $Y=(t-\tau_{\max})$ with travel times generated  with the one-sided L{\'e}vy distribution. For simulations, we use $10^5$ realizations, $\alpha=1/2$, and $v_0=1$. The blue circles describing the extreme events are predicted by Eq.~\eqref{18lwsec3106b}; see the inset.  The strong correlations show that the statistics of large positions are determined by the longest waiting times; see also Fig.~\ref{PositionVsTauMaxSimulation}.
}\label{levywalkTaumaxVSx}
\end{figure}

Here, we first define $$\tau_{\max}=\max\{\tau_1, \tau_2, \cdots, \tau_N, B_t\}.$$
The limiting law of typical fluctuations of $\tau_{\max}$ has been studied by C. Godr{\`{e}}che et al. in Ref.~\cite{Godreche2015Statistics}. Here we recently showed  that another law will be found when the second scaling $\tau_\text{max} \simeq t$ or $\tau_\text{max} \to t$ is introduced \cite{Marc2019}. If $\tau_{\max}\to t$, the density of $\tau_{\max}$ can be deduced from the following inverse Fourier transform
\begin{equation}\label{18lwsec5eq03op}
f_{\eta}(\eta,t)\sim \frac{b_\alpha}{ (t-\eta)^{\alpha}\Gamma(1-\alpha)}
\mathcal{F}^{-1}_{\eta}\left[ \frac{1}{1-\widehat{\phi}(-ik_\eta)}\right]
\end{equation}
with $\eta=t-\tau_{\max}$; see Eq.~\eqref{2019lgsec21c101} in Appendix \ref{19LGS6APPA}.  We  focus on the case where $\tau_{\max}$ is large and $t-\tau_{\max}\to 0$. The limit $k_\eta\to 0$ corresponds to the large `time' $\eta$.
Note that here we must consider the full form of the density, namely $\widehat{\phi}(- ik_\eta)$ is important, while for the typical fluctuations only the small $k_\eta$ behavior  of Eq.~\eqref{2019lgsec102} is important. 
Rewriting   Eq.~\eqref{18lwsec3105}, for the variable $\epsilon=v_0 t -x$ we have
\begin{equation}\label{18lwsec3105a}
\begin{split}
  P_{\epsilon}(\epsilon,t) &\sim\frac{b_\alpha}{ t^{\alpha}\Gamma(1-\alpha)}
\mathcal{F}^{-1}\left[\frac{1}{1-\widehat{\phi}(-2ik_\epsilon v_0)}\right] \\
    & =\frac{b_\alpha\Gamma(1-\alpha)^{-1}}{ 2v_0t^{\alpha}} \mathcal{F}^{-1}_{\epsilon/(2v_0)} \left[\frac{1}{1-\widehat{\phi}(-ik_\epsilon)}\right].
\end{split}
\end{equation}  
Combining Eqs.~\eqref{18lwsec5eq03op} and \eqref{18lwsec3105a}, we get the main result of this section
\begin{equation}\label{18lwsec3106b}
  t-\tau_{\max}\overset{d}{=} \frac{1}{2v_0}(v_0t-x),
\end{equation}
where $\tau_{\max}$ is large and $\overset{d}{=}$ means that the distributions of the random variables on both sides of Eq.~\eqref{18lwsec3106b}, i.e., $v_0t-x$ and $2v_0(t-\tau_{\max})$, are the same. Hence in Eq.~\eqref{18lwsec3106b} $x$ is  of course the position of the particle at large $t$, when it is in the vicinity of the `light' cone. 
As shown in Fig.~\ref{PositionVsTauMaxSimulation}, the density of $t-\tau_{\max}$ is consistent with that of $(t-x)/(2v_0)$ for large $x$ and $\tau_{\max}$.  Rewriting  Eq.~\eqref{18lwsec3106b} yields
\begin{equation}\label{18lwsec5eq03}
 x\overset{d}{=}v_0\tau_{\max}-v_0(t-\tau_{\max}).
\end{equation}
This behavior can  be tested  based on a correlation plot. As can be seen in Fig.~\ref{levywalkTaumaxVSx}, the strong relation between $t-\tau_{\max}$ and $(v_0t-x)/(2v_0)$ is illustrated \footnote{In our simulations we do not have the data below the line $X=Y$ since $x\geq \tau_{\max }v_0-(t-\tau_{\max}v_0)$ which yields $X\geq Y$.}. As expected, we find that  $t-\tau_{\max}$  grows linearly with $(v_0t-x)/(2v_0)$ for a small $t-\tau_{\max}$.

This relation stressed here is  different from  the case of $\alpha>1$ discussed in \cite{Alessandro2019Single}  in which the relation is $x\overset{d}{=}v_0\tau_{\max}$. The reason is as follows: For $\alpha>1$, the length of the displacement made in $(0,t-\tau_{\max})$, which is the time interval free of the longest waiting time, follows $x(t)\propto (t-\tau_{\max})^{1/\alpha}\ll t$. While, for $\alpha<1$, the situation is changed since $t-\tau_{\max}\propto t$ and the term $v_0(t-\tau_{\max})$ comes into play. See further details in Appendix \ref{19LGS6APPA2}.

We further treat Eq.~\eqref{18lwsec3106b} heuristically to explain its meaning. Now the total observation time $t$ is divided into two parts: One is the sum of waiting times denoted by $t_{{\rm positive}}$ when directions of particles are positive and  the other one is $t-t_{{\rm positive}}$.
We assume that $x\approx v_0t$, the particles arrive there by a mechanism of large jump. This means that we have $t_{{\rm positive}}=\tau_{\max}$ and $\tau_{\max} \approx t$. More specifically, we consider two random variables $\tau_{\max}$ and the remaining time $t-\tau_{\max}$. The corresponding position of the particle is $x(t)\approx v_0\tau_{\max}\pm v_0 (t-\tau_{\max})$.
We further suppose that the particle moves with velocity $+v_0$ in $(0,\tau_{\max})$ and $-v_0$ in the remaining time, then the position of the particle at time $t$ is
\begin{equation}\label{18lwsec5eq03f19}
  x\approx v_0\tau_{\max}-v_0(t-\tau_{\max}).
\end{equation}
If the particle does not change its direction in the mentioned two time intervals, we have $x(t)\approx v_0t$ (a delta function).

To conclude we see that rare events are obtained by a particle moving only in  one direction from Eq.~\eqref{18lwsec5eq03f19} (for $\tau_{\max}$) and reversing direction in the remaining time, we do not see this as an intuitive result, but when $\tau_{\max}$ is really big the particle  is left with little time to reverse, hence the ballistic motion with the reverse direction is plausible
(and if it continues in the same direction, we are on the horizon of the walk, which is not considered here). Note that in principle  the number of renewals related to Eq.~\eqref{18lwsec3106b} can be a large number and is not limited to one.
This will be discussed rigorously in the following section.

\section{Relation between rare events of the position and the average of renewals $\langle N\rangle$}\label{18lwsec5new}
\begin{figure}[htb]
 \centering
 \includegraphics[width=8cm]{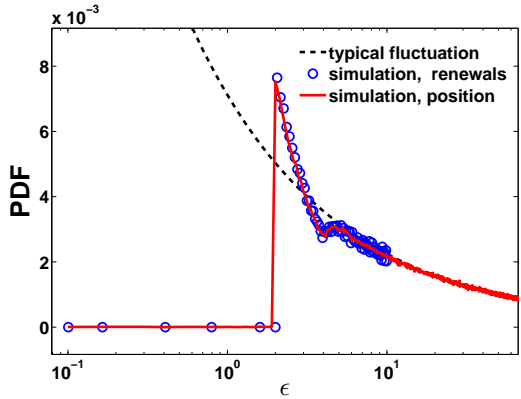}\\
 \caption{Simulations of the statistical behavior of $\epsilon=v_0t-x$ and  the derivative of the mean of the number of renewals predicted by Eq.~\eqref{18lwsec5new105}. The parameters are the same as in Fig.~\ref{PositionVsTauMaxSimulation}.
}\label{PositionVsRenewalSimulation}
\end{figure}

Now the aim is to investigate the relation between $x$ and the number of renewals. Note that the rare events of the position are governed by $\widehat{\phi}^n(-ik_\epsilon)$ with $n$ being a positive integer according to Eqs.~\eqref{18lwsec3106} and  \eqref{18lwsec3105a1d}. It indicates that the rare events of the position have a strong relationship with the number of renewals. Based on the renewal  theory \cite{Godreche2001Statistics}, in Laplace space ($t\to s$),
the probability  of the number of renewals during time interval $0$ and $t$ is \cite{Godreche2001Statistics}
\begin{equation}\label{18lwsec5new101}
\widehat{p}_N(s)=\widehat{\phi}^N(s)\frac{1-\widehat{\phi}(s)}{s}.
\end{equation}
We can check that $p_N(t)$ is normalized by using $\sum_{N=0}^\infty \widehat{p}_N(s)=1/s$. According to Eq.~\eqref{18lwsec5new101}, the  mean of renewals is
\begin{equation}\label{18lwsec5new102}
  \langle \widehat{N}(s)\rangle=\sum_{N=0}^\infty N\widehat{p}_N(s)=\frac{\widehat{\phi}(s)}{s(1-\widehat{\phi}(s))}.
\end{equation}
Rewriting Eq.~\eqref{18lwsec5new102}, we get
\begin{equation}\label{18lwsec5new103}
  s\langle \widehat{N}(s)\rangle=\frac{\widehat{\phi}(s)}{1-\widehat{\phi}(s)}=\frac{1}{1-\widehat{\phi}(s)}-1
\end{equation}
Taking the inverse Laplace transform yields
\begin{equation}\label{18lwsec5new104}
  \frac{d \langle N(t) \rangle}{d t}=\mathcal{L}^{-1}_t\left[\frac{1}{1-\widehat{\phi}(s)}\right]
\end{equation}
with $t>0$. Note that Eq.~\eqref{18lwsec5new104} is  the exact result for $t>0$ and $\mathcal{L}^{-1}_t[1/(1-\widehat{\phi}(s))$] corresponds to the rate of the number of renewals \cite{Takuma2020Infinite}.       Combining the second line of the right hand side of Eqs.~\eqref{18lwsec3105a} and \eqref{18lwsec5new104}, the relation between the behavior of $P_{\epsilon}(\epsilon,t)$ and $\langle N(t) \rangle$ is found
\begin{equation}\label{18lwsec5new105}
\begin{split}
  P_{\epsilon}(\epsilon,t) &\sim\frac{b_\alpha\Gamma(1-\alpha)^{-1}}{ 2v_0t^{\alpha}} \left(\frac{d \langle N(z) \rangle}{d z}\Bigg|_{z=\frac{\epsilon}{2v_0}}\right).
\end{split}
\end{equation}
\added{Eq.~\eqref{18lwsec5new105} is the main result of this section describing the relation between the position of the L{\'e}vy walker and the derivative of the average of the number of renewals at the value of $\epsilon/(2v_0)$.
Thus, we need to observe data only for a very short time and then we can map the observed number of renewals to the rare fluctuations of the position generated for an extremely long observation time $t$. This is  particularly important for real experiments to save a lot of time and expense since there is no need to record the data up to observation time $t$. In that sense we find an useful and important relation between positional rare fluctuations and 
the derivative of $\langle N\rangle$.}
Eq.~\eqref{18lwsec5new105} is illustrated  in Fig.~\ref{PositionVsRenewalSimulation}. 
Here the right-hand side of Eq.~\eqref{18lwsec5new105} is obtained by averaging $10^7$ realizations.
It can be seen that $P_{\epsilon}(\epsilon,t)$ is connected to the average of renewals at a small `time' $z=\epsilon/(2v_0)$. The simulations for the mean number of renewals were made only up to time $t=10$, still they predict the rare events with simulations made for time $t=1000$. 

\section{ Propagator for the Pareto traveling time PDF}\label{18lwsec5finiewqv1}
Now we deal with a general observation time $t$ instead of the long time limit considered in previous sections, then use a different method to explain the rare fluctuations again.  Here we focus on the case of  the waiting time following the Pareto distribution Eq.~\eqref{2019lgsec101}. Recall that $\tau_0$ is the cutoff for this distribution.
If $x\in (v_0t-\tau_0v_0,v_0t)$, from Eq.~\eqref{18lwsec310600} we get $P(x,t)= 0$.
Indeed in Fig. \ref{LevywalkVeolicity05Large} we see this effect rather easily, however this is an approximation valid in the long time limit only, for finite times this rule is not strictly valid as the probability of finding the particle in this interval is not identically zero.
Intuitively, the large position is related to the large waiting time and the large waiting time is determined by the far tail of the waiting time. So it would be interesting to consider the relationship between the far tail of the waiting time and the large position.
%
%
%
%
%
%

\begin{figure}[htb]
 \centering
 \includegraphics[width=8cm, height=6cm]{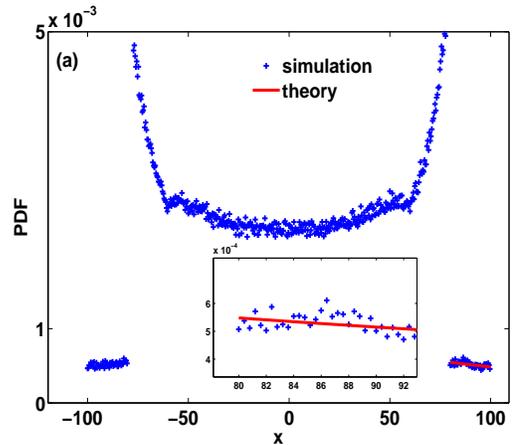}\\
 \caption{The PDF of the position with a finite observation time $t$. The solid line is the theoretical prediction obtained from  Eq.~\eqref{18lwsec5finiev5}, showing derivations from the rare fluctuations Eq.~\eqref{18lwsec310600}; see the inset.
Here we choose $\alpha=1/2$, $\tau_0=10$, $N=10^6$, $t=100$.
}\label{FiniteTime}
\end{figure}

Using Taylor's expansion on Eq.~\eqref{18lwsec3101}, we get
\begin{equation}\label{18lwsec5finiev2}
\begin{split}
 \widetilde{\widehat{P}}(k,s)  =&\left(\widehat{\Psi}(s+ikv_0)+\widehat{\Psi}(s-ikv_0)\right)/2\\
    & \times \sum_{n=0}^{\infty} \left(\frac{\widehat{\phi}(s+ikv_0)+\widehat{\phi}(s-ikv_0)}{2}\right)^n.
\end{split}
\end{equation}
Here the summation over $n$ is a sum over the number of renewals, which as mentioned is random. We focus on the case of $x\in (v_0t-2v_0\tau_0,v_0t)$. If $n=0$, then clearly the particle is not in the interval under study. In order to obtain $P(x,t)$ in the mentioned spatial interval, we need to consider the propagator Eq.~\eqref{18lwsec5finiev2}.  Utilizing the definition of $\phi(x,t)$ and $\Psi(x,t)$, for $x\in (v_0t-2v_0\tau_0,v_0t)$ the above equation reduces to
\begin{equation}\label{18lwsec5finiev3}
\begin{split}
\widetilde{\widehat{P}}(k,s)  =&\frac{1}{2}\widehat{\Psi}(s+ikv_0) \sum_{n=1}^{\infty} \left(\frac{\widehat{\phi}(s-ikv_0)}{2}\right)^n.\\
\end{split}
\end{equation}
The infinite terms on the right-hand side of Eq.~\eqref{18lwsec5finiev3} describe the probability of moving in  positive direction all the time but the direction of the last step is negative. This is the only way the particles can reach $ (v_0t-2v_0\tau_0,v_0t)$ at time $t$.
Taking the inverse Laplace-Fourier transform, we obtain
\begin{equation}\label{18lwsec5finiev4}
\begin{split}
P(x,t) & =\frac{1}{4v_0}\int_{\frac{t-x/v_0}{2}}^{\infty}\phi(y)dy \mathcal{ L}^{-1}_{\frac{t+x/v_0}{2}}\left[\frac{\widehat{\phi}(s)}{2-\widehat{\phi}(s)}\right], \\
\end{split}
\end{equation}
which reduces to
\begin{equation}\label{18lwsec5finiev5}
\begin{split}
P(x,t) &=\frac{1}{4v_0}\left( \mathcal{L}^{-1}_{\frac{t+x/v_0}{2}}\left[\frac{\widehat{\phi}(s)}{2-\widehat{\phi}(s)}\right]\right), \\
\end{split}
\end{equation}
with $x\in (v_0t-2v_0\tau_0,v_0t)$.  When $t$ is large, i.e., $\widehat{\phi}(s)\sim 1-b_\alpha s^\alpha$, we have
\begin{equation}\label{18lwsec5finiev5fjjdh}
\begin{split}
P(x,t) &=\frac{1}{4v_0}\Big( \mathcal{L}^{-1}_{\frac{t+x/v_0}{2}}\Big[
\frac{1-b_\alpha s^\alpha}{1+b_\alpha s^\alpha}\Big]\Big)\\
&\propto -\mathcal{L}^{-1}_{\frac{t+x/v_0}{2}} [s^\alpha].
\end{split}
\end{equation}
Thus,
\begin{equation}\label{18lwsec5finiev5apps}
\begin{split}
P(x,t) &\propto \left(\frac{t}{2}+\frac{x}{2v_0}\right)^{-\alpha-1}.
\end{split}
\end{equation}
Clearly, the far tail of the position decays as a power law.
Contrary to Eq.~\eqref{18lwsec310600} with $t\to \infty$, the behavior of $x\in (v_0t-2\tau_0v_0,v_0t)$ is governed by the far tail of the PDF for a finite time; see Fig.~\ref{FiniteTime}. With the increasing of observation time $t$, Eq.~\eqref{18lwsec5finiev5} goes to zero and approaches Eq.~\eqref{18lwsec310600} since  the probability of reaching $(v_0t-2\tau_0v_0,v_0t)$ becomes smaller and smaller.

We proceed with the discussion of $x=v_0t$. Recall that for Eq.~\eqref{18lwsec3106b}, it is meaningless for $\tau_{\max}=t$. When all directions of the particles are the same, this  contributes to the probability of $x=v_0t$. For this maximum point, we have
\begin{equation}\label{18lwsec5finiev6}
\begin{split}
\widetilde{\widehat{P}}(k,s)  =&\frac{1}{2}\widehat{\Psi}(s-ikv_0) \sum_{n=0}^{\infty} \left(\frac{\widehat{\phi}(s-ikv_0)}{2}\right)^n
\end{split}
\end{equation}
with $x=v_0t$.  The inversion of Eq.~\eqref{18lwsec5finiev6} is
\begin{equation}\label{18lwsec5finiev7}
\begin{split}
  \widehat{P}(x,t)  =\frac{1}{2}\delta(x-v_0t)\mathcal{L}^{-1}_t\left[\frac{\widehat{\Psi}(s)}{1-\widehat{\phi}(s)/2}\right].
\end{split}
\end{equation}
In the long time limit, using $1-\widehat{\phi}(s)/2\sim 1/2$, Eq.~\eqref{18lwsec5finiev7} reduces to
\begin{equation}\label{18lwsec5finiev8}
\begin{split}
  \widehat{P}(x,t)  \sim\delta(x-v_0t)\Psi(t).
\end{split}
\end{equation}
It can be seen that Eq.~\eqref{18lwsec5finiev8}  depends on the initial position of the particle and is related to the survival probability.
In reality, Eq.~\eqref{18lwsec5finiev8} is related to the single big jump principle, namely, the particles go in the one direction for the first step and continue in the same way for the rest steps.

In summary: We have carried out an investigation on the moving forward particles, leading to a delta function contribution at $x=v_0t$; see Eq.~\eqref{18lwsec5finiev8}. This is clearly a description of a rare event, hence the theory is developed in two stages, $x$ large and  comparable to $v_0t$ but strictly smaller, and $x=v_0t$.
Note that when $x=v_0 t$, the particle did not change its direction, but the renewal process may have many collision events. If the velocity distribution is not $+v_0$ and $-v_0 $ with equal probability we might obtain different behaviors than what we presented here, however this is hardly surprising as it is also true  for the bulk \cite{Froemberg2013Random}.

One may wonder how can we  understand the mechanism of rare fluctuations discussed in previous section. Motivated by the finite time limit Eq.~\eqref{18lwsec5finiev5apps}, we consider the case that  the particle just has only one negative velocity (the rest epochs are positive) to study the behavior of $x\to v_0t$. Similar to Eq.~\eqref{18lwsec5finiev3}, we obtain from Eq.~\eqref{18lwsec5finiev2}
\begin{equation}\label{18lwsec5finievseew12}
\begin{split}
\widetilde{\widehat{P}}(k,s) & \sim\frac{\widehat{\phi}(s+ikv_0)}{2}\frac{\widehat{\Psi}(s-ikv_0)}{2}\sum_{n=0}^\infty \left(\frac{\widehat{\phi}(s-ikv_0)}{2}\right)^2\\
    &~~~+\frac{\widehat{\Psi}(s+ikv_0)}{2}\left(\sum_{n=1}^\infty \left(\frac{\widehat{\phi}(s-ikv_0)}{2}\right)^2 \right).
\end{split}
\end{equation}
Here the second line of the right hand side of  Eq.~\eqref{18lwsec5finievseew12} corresponds to case where the negative velocity is only in the last step and the first line is when the  sole negative direction is not the last one.
The sum of the infinite  geometric  series yields
\begin{equation}\label{18lwsec5finievseew13}
\begin{split}
\widetilde{\widehat{P}}(k,s) & \sim\frac{\widehat{\phi}(s+ikv_0)}{2}\frac{\widehat{\Psi}(s-ikv_0)}{2}\frac{1}{1-\frac{\widehat{\phi}(s-ikv_0)}{2}}
  \\
    &~~~+\frac{\widehat{\Psi}(s+ikv_0)}{2}\frac{1}{1-\frac{\widehat{\phi}(s-ikv_0)}{2}}.
\end{split}
\end{equation}
Using relation Eq.~\eqref{18lwsec3102} again, we have
\begin{equation}\label{18lwsec5finievseew13fg1}
\begin{split}
\widetilde{\widehat{P}}_{\epsilon}(k_\epsilon,s) & \sim\frac{\widehat{\phi}(s-2ik_\epsilon v_0)}{2}\frac{\widehat{\Psi}(s)}{2}\frac{1}{1-\frac{\widehat{\phi}(s)}{2}}
  \\
    &~~~+\frac{\widehat{\Psi}(s-2ik_\epsilon v_0)}{2}\frac{1}{1-\frac{\widehat{\phi}(s)}{2}}.
\end{split}
\end{equation}
The inverse Laplace transform of the above equation gives
\begin{equation}\label{18lwsec5finievseew13fg1sda}
\begin{split}
P_{\epsilon}(\epsilon,s) & \sim \frac{\phi(\frac{\epsilon}{2v_0})}{2v_0}\frac{b_\alpha}{\Gamma(1-\alpha)t^\alpha}+\frac{1}{4v_0}\phi\left(t-\frac{\epsilon}{2v_0}\right).
\end{split}
\end{equation}
Based on Eq.~\eqref{18lwsec5finievseew13fg1sda}, the leading term of Eq.~\eqref{18lwsec3106} is obtained again in the limit $t \to \infty$.
It can be seen that Eq.~\eqref{18lwsec5finievseew13fg1} or \eqref{18lwsec5finievseew13fg1sda} is an exact solution with Eq.~\eqref{2019lgsec101} when $x\in (v_0t-3\tau_0v_0,v_0t-2\tau_0v_0)$. Besides, it is easy to find that the rare events of the position are closely contact to the number of the negative directions. This is also corresponding to the expanded terms of Eq.~\eqref{18lwsec3105a1} in powers of $\widetilde{\phi}(-ik_{\epsilon})$.


\section{Conclusion}\label{18lwsec6}

The main focus of this manuscript has been on the rare fluctuations of the ballistic  L{\'e}vy walk model  in one dimension. \added{We show here that the density $P(x,t)$ in the vicinity of the ballistic cone $x \simeq v_0 t$, is described by the full shape of the distribution of waiting times, unlike the typical fluctuations which are described by the Lamperti-arcsine law.}  To highlight the rare fluctuations, we use a second non ballistic scaling. Namely, we multiply the PDF $P(\epsilon,t)$ with $t^\alpha$ and obtain the infinite density $\mathcal{I}(\epsilon)$; see Eqs.~\eqref{18lwsec3105a1} and \eqref{18lwsec3105}. The integral of $\mathcal{I}(\epsilon)$ diverges at large $\epsilon=v_0t-x$ \added{and in that sense it is a non-normalizable solution which is hardly surprising.} The infinite density and the normalized Lamperti-arcsine are complementary, with the former describing the positional distribution in the vicinity of the `light' cone, namely close to the maximum of the packet.  Certainly, our infinite density is different from the case of $1<\alpha<2$ \cite{Rebenshtok2014Non}. 
We gave the relation between the infinite density of the \added{maximum of} waiting times and the infinite density of the position $x$. \added{In this sense we relate between extreme value statistic of constrained random variables ($\tau_{\max}$) and the position of the L{\'e}vy walk close to the ballistic  cone. This gives a single big jump principle for the ballistic L{\'e}vy walk Eq.~\eqref{18lwsec5eq03}. We note that this formula differs from the one found in the super-diffusive phase of the L{\'e}vy walk model} \cite{Alessandro2019Single}. 
We also analysed the moving forward particles which lead to a delta function contribution [see Eq.~\eqref{18lwsec5finiev8}] and are also clearly rare events in the long time limit.

We have developed a rate formalism to the rare events, see Eq. \eqref{18lwsec5new105}. Recently, Akimoto et al. considered a related rate formalism for a different observable: the velocity \cite{Takuma2020Infinite}, while here we consider the position. Indeed the rate approach is a valuable  tool for the calculations of large fluctuations, at least for renewal processes. This rate formalism was initially developed by Vezzani et al  \cite{Alessandro2019Single,Vezzani2020Rare}. In real experiments \added{or simulations with trajectories generated with a computer program}, the observation of the rare fluctuations  is a  challenge, since we need many samples and a  long observation time $t$. Here we investigate the rare fluctuations of the position from a new and different point of view.
Utilizing the renewal theory, we find that the rare events of the L\'{e}vy walk are related to the mean number of renewals in the observation time $(0, (x-v_0t)/(2v_0))$. This is to say, if we are interested in the rare events of the position of the ballistic L{\'e}vy walk  model, we just need the data of the renewals at some  `time' $(x-v_0t)/(2v_0))$; see Eq.~\eqref{18lwsec5new105}. \added{And since $v_0 t$ is of order $x$, the `time' here is actually very short, so the knowledge of the mean number of renewals for short times provides the full information on the rare events for large times, which is remarkable.}



\section*{Acknowledgments}

E. B. acknowledges the Israel Science Foundation for support through  Grant No. 1898/17. M.H. is funded by the Deutsche Forschungsgemeinschaft (DFG, German Research Foundation) - 436344834.
W.W. thanks Felix Thiel for the discussions.
W.W. was supported by Bar-Ilan University together with the Planning and Budgeting Committee fellowship program.

\appendix
\begin{appendices}

\section{Calculation of the PDF of $\tau_{\max}$
}\label{19LGS6APPA}
Let us  give a brief account of the statistic of $\tau_{\max}=\max\{\tau_1,\tau_2,\cdots,\tau_N,B_t\}$. In Laplace space, the density of $\tau_{\max}$ satisfies \cite{Godreche2015Statistics}
\begin{equation}\label{2019lgsec2a01}
\int_{M}^{\infty} \widehat{f}_{\tau_{\max}}(z,s)dz=\frac{1}{s}\frac{1}{1+\widehat{g}(M,s)}
\end{equation}
with
\begin{equation}\label{2019lgsec2a02}
    \widehat{g}(M,s)=\frac{s\exp(sM)\int_0^{M} \Psi(z)\exp(-sz)dz }{\int_{M}^{\infty}\phi(z)dz}.
\end{equation}
From Eq.~\eqref{2019lgsec2a01}, the density of $\tau_{\max}$ follows
\begin{equation}\label{2019lgsec21c101w}
\widehat{f}_{\tau_{\max}}(M,t)\sim \frac{\exp(-sM)}{1-\widehat{\phi}(s)}
\left(\Psi(M)+\frac{\phi(M)}{s}\right).
\end{equation}
In the limit $t\to \infty$ and $M\to t$,   the leading term of Eq.~\eqref{2019lgsec21c101w} is
\begin{equation}\label{2019lgsec21c101}
\begin{split}
 \widehat{f}_{\tau_{\max}}(M,s) & \sim \frac{\Psi(M)\exp(-sM)}{1-\widehat{\phi}(s)}.
\end{split}
\end{equation}
Notice that Eq.~\eqref{2019lgsec21c101} can be further simplified. Substituting  Eq.~\eqref{2019lgsec102} into Eq.~\eqref{2019lgsec21c101} and taking the inverse Laplace transform give
\begin{equation}\label{2019lgsec21b104}
 f_{\tau_{\max}}(M,t)\sim \frac{\sin(\pi\alpha)}{\pi}(t-M)^{\alpha-1}t^{-\alpha}.
\end{equation}
Rewriting Eq.~\eqref{2019lgsec21b104} yields the scaling form of $f_{\tau_{\max}}(M,t)$
\begin{equation}\label{2019lgsec21b105}
    f_{x=\tau_{\max}/t}(x)\sim \frac{\sin(\pi\alpha)}{\pi}(1-x)^{\alpha-1};
\end{equation}
see also \cite{Godreche2015Statistics}. Let us now take the integral over $x$ from $0$ to $1$ of Eq.~ \eqref{2019lgsec21b105}, we get
\begin{equation}\label{2019lgsec21b10add1}
\int_0^1f_{x=\tau_{\max}/t}(x)dx\sim\frac{\sin(\pi\alpha)}{\pi\alpha}.
\end{equation}
It indicates that only $\alpha\to 0$, the density is normalized otherwise not. Note that Eq.~\eqref{2019lgsec21b104} is valid under condition that $t\to \infty$ and $t- \tau_{\max}$ is large . While, when another scaling is introduced, i.e., $\tau_{\max}\to t$, the statistics of $\tau_{\max}$ will be changed. Using the definition of inverse Laplace transform on Eq.~\eqref{2019lgsec21c101} gives
\begin{equation}\label{2019lgsec21b107}
f_{\tau_{\max}}(M,t)\sim \frac{\Psi(M)}{2\pi i}\int_{a-i\infty}^{a+i\infty} \frac{\exp(s(t-M))}{1-\widehat{\phi}(s)}ds.
\end{equation}
When $s\to 0$, the random variable $t-M$ tends to infinity and in this limit Eq.~\eqref{2019lgsec21b107} corresponds to the statistics of large $t-M$. While,  for  $t-M\to 0$ we need the information of large $s$ rather than $s\to 0$. In other words, the dynamic of $M\to t$ is governed by the full form of $\phi(\tau)$.
Here we just  use the density of $\tau_{\max}$ to build the relationship between $\tau_{\max}$ and $x$,
and discussion of Eq.~\eqref{2019lgsec21c101w} or \eqref{2019lgsec21c101} will be shown by another paper \cite{Marc2019}.

\begin{widetext}
\begin{figure*}[htb]
 \centering
 \includegraphics[width=14cm, height=8cm]{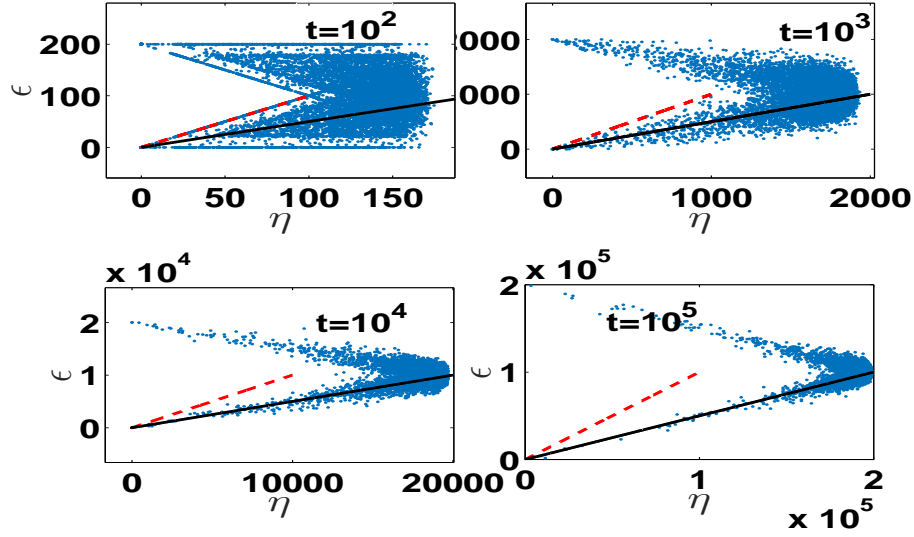}\\
 \caption{Correlation plot between $\eta=2v_0(t-\tau_{\max})$ and $\epsilon=v_0t-x$ shown in Eq.~\eqref{18lwsec3106b} for a L\'{e}vy walk model with different observation time $t$.   When the observation time $t$ is not very large, for example $t=10^2$, the statistics of the single big jump Eq.~\eqref{18lwsec3106b} shown by the dashed red line are found.  On the other hand, if $t$ is enough large, the distribution of the biggest position and the largest waiting times agree with each other; see the black solid line. As mentioned this is a diversion, since $\alpha>1$. Note that the slope of the red dashed and the black solid lines are one and half, respectively.
The parameters are the same as in Fig.~\ref{LevyWalk15distribution}.
}\label{LevyWalk15Scatter}
\end{figure*}
\end{widetext}

\begin{figure}[htb]
 \centering
 \includegraphics[width=8cm, height=6cm]{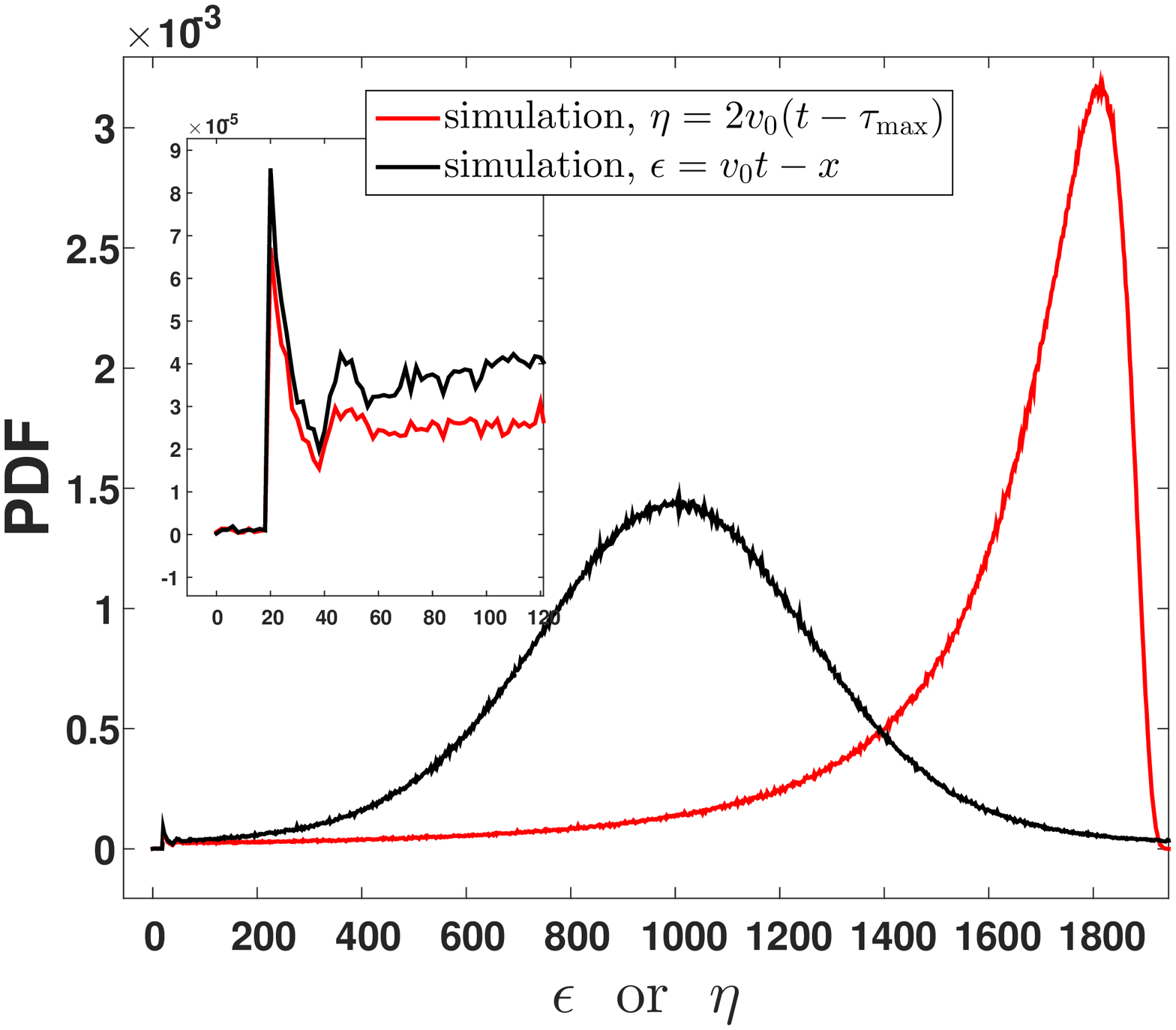}\\
 \caption{The PDFs of random variables $\eta=2v_0(t-\tau_{\max})$ and $\epsilon=v_0t-x$  with $\alpha=1.5$ for a L\'{e}vy walk model. We use the Pareto PDF for the travel times, and choose
$t=1000$, $v_0=1$, $\tau_0=10$, and $N=6\times10^6$. When $\epsilon$ and $\eta$ are small, clearly the distribution of $\eta$ is the same as that of $\epsilon$; see the inset.
}\label{LevyWalk15distribution}
\end{figure}

\section{The relation between the largest position and the longest traveling time with $0<\alpha<2$}\label{19LGS6APPA2}
Here we make a brief comparison for the rare events of position and the waiting time.
Contrary to $1<\alpha<2$, the rare event of the position is not only determined by the longest waiting  times but also by another fluctuant  term. Though the micro term is small, we can not ignore it; see Eq.~\eqref{18lwsec5eq03}. This relation can also be found for the case of $\alpha>1$; see Figs.~\ref{LevyWalk15distribution} and \ref{LevyWalk15Scatter} from the view of the distribution and the correlation plot. This means that in some special region near $v_0 t$ the rare fluctuations are determined by the full form of the waiting time even for $1<\alpha<2$. While, with the decrease of $\tau_{\max}$, the behavior is governed by the asymptotic behavior of the distribution of the waiting time showing the behavior when $v_0\tau_{\max}\propto x$ but $v_0\tau_{\max}\not \approx x$; see Ref. \cite{Alessandro2019Single}. As expected, the single big jump principle Eq.~\eqref{18lwsec3106b} under study, will  vanish due to $\tau_0/t\to 0$ with the increase of $t$ and $1<\alpha<2$. With the help of our work, one can get a better understanding about the rare events, the relation between $0<\alpha<1$ and $1<\alpha<2$, and the infinite densities.

\end{appendices}

\bibliographystyle{prestyle}
\bibliography{wenxian}

\begin{thebibliography}{10}

\bibitem{Einstein1905Uber}
A.~Einstein, Ann. Phys. \textbf{322}, 549 (1905).

\bibitem{Smoluchowski1906Zur}
M.~von Smoluchowski, Ann. Phys. \textbf{326}, 756 (1906).

\bibitem{Metzler2000random}
R.~Metzler and J.~Klafter, Phys. Rep. \textbf{339}, 1 (2000).

\bibitem{Davydov1934Uravneniya}
D.~B. I, DoM. Akad. Nauk SSSR \textbf{2}, 474 (1934).

\bibitem{Goldstein1951diffusion}
S.~Goldstein, Q. J. Mech. Appl. Math. \textbf{4}, 129 (1951).

\bibitem{Joseph1989Heat}
D.~D. Joseph and L.~Preziosi, Rev. Mod. Phys. \textbf{61}, 41 (1989).

\bibitem{Masoliver1996Finite}
J.~Masoliver and G.~H. Weiss, Eur. J. Phys \textbf{17}, 190 (1996).

\bibitem{Weiss2002Some}
G.~H. Weiss, Physica A \textbf{311}, 381  (2002).

\bibitem{Bakunin2003Mysteries}
O.~G. Bakunin, Phys. Usp. \textbf{46}, 309 (2003).

\bibitem{Masoliver2017Continuous}
J.~Masoliver and K.~Lindenberg, Eur. Phys. J. B \textbf{90}, 1 (2017).

\bibitem{Zaburdaev2015Levy}
V.~Zaburdaev, S.~Denisov, and J.~Klafter, Rev. Mod. Phys. \textbf{87}, 483
  (2015).

\bibitem{Compte1997generalized}
A.~Compte and R.~Metzler, J. Phys. A: Math. Gen. \textbf{30}, 7277 (1997).

\bibitem{Eckstein1999mathematics}
E.~C. Eckstein, J.~A. Goldstein, and M.~Leggas, Electron. J. Differential
  Equations \textbf{1999}, 39 (1999).

\bibitem{Cascaval2002Fractional}
R.~C. Cascaval, E.~C. Eckstein, C.~L. Frota, and J.~A. Goldstein, J. Math.
  Anal. Appl. \textbf{276}, 145 (2002).

\bibitem{Qi2011Solutions}
H.~Qi and X.~Jiang, Physica A \textbf{390}, 1876  (2011).

\bibitem{Fedotov2016Single}
S.~Fedotov, Phys. Rev. E \textbf{93}, 020101(R) (2016).

\bibitem{Awad2019On}
E.~Awad, Physica A \textbf{518}, 210 (2019).

\bibitem{Zumofen1993Scale}
G.~Zumofen and J.~Klafter, Phys. Rev. E \textbf{47}, 851 (1993).

\bibitem{Fogedby1994Levy}
H.~C. Fogedby, Phys. Rev. Lett. \textbf{73}, 2517 (1994).

\bibitem{Shlesinger1995Levy}
M.~F. Shlesinger, G.~M. Zaslavsky, and U.~Frisch (eds.), \emph{L\'{e}vy flights
  and related topics in physics}, vol. 450 of \emph{Lecture Notes in Physics}
  (Springer-Verlag, Berlin, 1995).

\bibitem{Kessler2012Theory}
D.~A. Kessler and E.~Barkai, Phys. Rev. Lett. \textbf{108}, 230602 (2012).

\bibitem{Shlesinger1982Random}
M.~F. Shlesinger, J.~Klafter, and Y.~M. Wong, J. Stat. Phys. \textbf{27}, 499
  (1982).

\bibitem{Klafter1994Levy}
J.~Klafter and G.~Zumofen, Phys. Rev. E \textbf{49}, 4873 (1994).

\bibitem{Zaburdaev2011Perturbation}
V.~Zaburdaev, S.~Denisov, and P.~H\"anggi, Phys. Rev. Lett. \textbf{106},
  180601 (2011).

\bibitem{Zaburdaev2013Space}
V.~Zaburdaev, S.~Denisov, and P.~H\"anggi, Phys. Rev. Lett. \textbf{110},
  170604 (2013).

\bibitem{Ramos2004Levy}
G.~Ramos-Fern{\'a}ndez, J.~L. Mateos, O.~Miramontes, G.~Cocho, H.~Larralde, and
  B.~Ayala-Orozco, Behav. Ecol. Sociobiol. \textbf{55}, 223 (2004).

\bibitem{Ariel2015Swarming}
G.~Ariel, A.~Rabani, S.~Benisty, J.~D. Partridge, R.~M. Harshey, and A.~Be'er,
  Nat. Commun. \textbf{6}, 8396 (2015).

\bibitem{Magdziarz2016Explicit}
M.~Magdziarz and T.~Zorawik, Phys. Rev. E \textbf{94}, 022130 (2016).

\bibitem{Zaburdaev2016Superdiffusive}
V.~Zaburdaev, I.~Fouxon, S.~Denisov, and E.~Barkai, Phys. Rev. Lett.
  \textbf{117}, 270601 (2016).

\bibitem{Marcin2017Aging}
M.~Magdziarz and T.~Zorawik, Phys. Rev. E \textbf{95}, 022126 (2017).

\bibitem{Fouxon2017Limit}
I.~Fouxon, S.~Denisov, V.~Zaburdaev, and E.~Barkai, J. Phys. A: Math. Theor.
  \textbf{50}, 154002 (2017).

\bibitem{Song2018Neuronal}
M.~S. Song, H.~C. Moon, J.-H. Jeon, and H.~Y. Park, Nat. Commun. \textbf{9}, 1
  (2018).

\bibitem{Giona2019Age}
M.~Giona, M.~D'Ovidio, D.~Cocco, A.~Cairoli, and R.~Klages, J. Phys. A: Math.
  Theor. \textbf{52}, 384001 (2019).

\bibitem{Bologna2020Distribution}
M.~Bologna, J. Stat. Mech: Theory Exp. \textbf{2020}, 073201 (2020).

\bibitem{Extended2020Giona}
M.~Giona, A.~Cairoli, and R.~Klages, ArXiv:2009.13434.

\bibitem{Kanazawa2020Loopy}
K.~Kanazawa, T.~G. Sano, A.~Cairoli, and A.~Baule, Nature \textbf{579}, 364
  (2020).

\bibitem{Margolin2005Nonergodicity}
G.~Margolin and E.~Barkai, Phys. Rev. Lett. \textbf{94}, 080601 (2005).

\bibitem{Froemberg2015Asymptotic}
D.~Froemberg, M.~Schmiedeberg, E.~Barkai, and V.~Zaburdaev, Phys. Rev. E
  \textbf{91}, 022131 (2015).

\bibitem{Aaronson1997introduction}
J.~Aaronson, \emph{An {I}ntroduction to {I}nfinite {E}rgodic {T}heory}, vol.~50
  of \emph{Mathematical Surveys and Monographs} (American Mathematical Society,
  Providence, RI, 1997).

\bibitem{Rebenshtok2014Infinite}
A.~Rebenshtok, S.~Denisov, P.~H\"anggi, and E.~Barkai, Phys. Rev. E
  \textbf{90}, 062135 (2014).

\bibitem{Rebenshtok2014Non}
A.~Rebenshtok, S.~Denisov, P.~H\"anggi, and E.~Barkai, Phys. Rev. Lett.
  \textbf{112}, 110601 (2014).

\bibitem{Erez2017Large}
E.~Aghion, D.~A. Kessler, and E.~Barkai, Phys. Rev. Lett. \textbf{118}, 260601
  (2017).

\bibitem{Wang2018Renewal}
W.~Wang, J.~H.~P. Schulz, W.~H. Deng, and E.~Barkai, Phys. Rev. E \textbf{98},
  042139 (2018).

\bibitem{Erez2019From}
E.~Aghion, D.~A. Kessler, and E.~Barkai, Phys. Rev. Lett. \textbf{122}, 010601
  (2019).

\bibitem{Wang2019Ergodic}
X.~Wang, W.~Deng, and Y.~Chen, J. Chem. Phys. \textbf{150}, 164121 (2019).

\bibitem{Cistjakov1964theorem}
V.~P. Chistjakov, Theory Probab. Appl. \textbf{9}, 640 (1964).

\bibitem{Alessandro2019Single}
A.~Vezzani, E.~Barkai, and R.~Burioni, Phys. Rev. E \textbf{100}, 012108
  (2019).

\bibitem{Wang2019Transport}
W.~Wang, A.~Vezzani, R.~Burioni, and E.~Barkai, Phys. Rev. Res. \textbf{1},
  033172 (2019).

\bibitem{Vezzani2020Rare}
A.~Vezzani, E.~Barkai, and R.~Burioni, Sci. Rep. \textbf{10}, 1 (2020).

\bibitem{Feller1971introduction}
W.~Feller, \emph{An {I}ntroduction to {P}robability {T}heory and {I}ts
  {A}pplications. {V}ol. {II}}.
\newblock Second edition (John Wiley \& Sons, Inc., New York, 1971).

\bibitem{Godreche2001Statistics}
C.~Godr{\`e}che and J.~M. Luck, J. Stat. Phys. \textbf{104}, 489 (2001).

\bibitem{Godreche2015Statistics}
C.~Godr\`eche, S.~N. Majumdar, and G.~Schehr, J. Stat. Mech: Theory Exp.
  \textbf{2015}, P03014 (2015).

\bibitem{Klafter1987Stochastic}
J.~Klafter, A.~Blumen, and M.~F. Shlesinger, Phys. Rev. A \textbf{35}, 3081
  (1987).

\bibitem{Peter1990Reaction}
P.~H\"anggi, P.~Talkner, and M.~Borkovec, Rev. Mod. Phys. \textbf{62}, 251
  (1990).

\bibitem{Bouchaud1990Anomalous}
J.-P. Bouchaud and A.~Georges, Phys. Rep. \textbf{195}, 127 (1990).

\bibitem{Fernando2009Beyond}
F.~D. Stefani, J.~P. Hoogenboom, and E.~Barkai, Phys. Today \textbf{62}, 34
  (2009).

\bibitem{Tameem2017Temperature}
T.~Albash, V.~Martin-Mayor, and I.~Hen, Phys. Rev. Lett. \textbf{119}, 110502
  (2017).

\bibitem{Klafter2011First}
J.~Klafter and I.~M. Sokolov, \emph{First Steps in Random Walks: From Tools to
  Applications} (Oxford University Press, Oxford, 2011).

\bibitem{Podlubny1999Fractional}
I.~Podlubny, \emph{Fractional Differential Equations} (Academic Press, Inc.,
  San Diego, 1999).

\bibitem{Kozubowski2001Fractional}
T.~J. Kozubowski, Math. Comput. Modelling \textbf{34}, 1023 (2001).

\bibitem{Vainstein2006Non}
M.~H. Vainstein, I.~V.~L. Costa, R.~Morgado, and F.~A. Oliveira, EPL
  \textbf{73}, 726 (2006).

\bibitem{Oliveira2019Anomalous}
F.~A. Oliveira, R.~M.~S. Ferreira, L.~C. Lapas, and M.~H. Vainstein, Front.
  Phys \textbf{7}, 18 (2019).

\bibitem{Lamperti1958occupation}
J.~Lamperti, Trans. Amer. Math. Soc. \textbf{88}, 380 (1958).

\bibitem{Gumbel2004Statistics}
E.~J. Gumbel, \emph{Statistics of {E}xtremes} (Dover Publications, Inc.,
  Mineola, 2004).

\bibitem{Albeverio2006Extreme}
S.~Albeverio, V.~Jentsch, and H.~Kantz (eds.), \emph{Extreme {E}vents in
  {N}ature and {S}ociety} (Springer, Berlin, 2006).

\bibitem{Majumdar2009Large}
S.~N. Majumdar and M.~Vergassola, Phys. Rev. Lett. \textbf{102}, 060601 (2009).

\bibitem{Lefevere2011Large}
R.~Lefevere, M.~Mariani, and L.~Zambotti, Stoch. Process. Their Appl.
  \textbf{121}, 2243 (2011).

\bibitem{Tsirelson2013uniform}
B.~Tsirelson, Electron. Comm. Probab. \textbf{18} (2013).

\bibitem{Krapivsky2014Large}
P.~L. Krapivsky, K.~Mallick, and T.~Sadhu, Phys. Rev. Lett. \textbf{113},
  078101 (2014).

\bibitem{Chaitra2014Universal}
C.~Hegde, S.~Sabhapandit, and A.~Dhar, Phys. Rev. Lett. \textbf{113}, 120601
  (2014).

\bibitem{Mauro2016Large}
M.~Mariani and L.~Zambotti, Adv. in Appl. Probab. \textbf{48}, 648 (2016).

\bibitem{Hartich2019Extreme}
D.~Hartich and A.~Godec, J. Phys. A \textbf{52}, 244001 (2019).

\bibitem{Marc2019}
M.~H\"{o}ll, W.~L. Wang, and E.~Barkai, ArXiv:2006.06253.

\bibitem{Thomas2013Precise}
T.~Mikosch and O.~Wintenberger, Probab. Theory Related Fields \textbf{156}, 851
  (2013).

\bibitem{Buraczewski2013Large}
D.~Buraczewski, E.~Damek, T.~Mikosch, and J.~Zienkiewicz, Ann. Appl. Probab.
  \textbf{41}, 2755 (2013).

\bibitem{Takuma2020Infinite}
T.~Akimoto, E.~Barkai, and G.~Radons, Phys. Rev. E \textbf{101}, 052112 (2020).

\bibitem{Froemberg2013Random}
D.~Froemberg and E.~Barkai, Eur. Phys. J. B \textbf{86}, 331 (2013).

\end{thebibliography}

\end{document}